# Smoothing representation of fitness landscapes – the genotype-phenotype map of evolution

Torsten Aßelmeyer, Werner Ebeling, Helge Rosé
*Institut für Physik, Humboldt–Universität zu Berlin,*
*Invalidenstraße 110, D-10115 Berlin, Germany*
(April 20, 1995)

We investigate an simple evolutionary game of sequences and demonstrate on this example the structure of fitness landscapes in discrete problems. We show the smoothing action of the genotype-phenotype mapping which still makes it feasible for evolution to work. Further we propose the density of sequence states as a classifying measure of fitness landscapes.

## I. INTRODUCTION

The fundamental dynamical processes of evolution are connected with dynamical processes based on sequences. This statement is supported by the basic Darwinian mechanism of molecular evolution. The genetic message - the *genotype* - is coded by a DNA-sequence and the whole cell dynamics is determined by an interplay of proteins and polynucleotides.

The *phenotype* are all properties characterizing a species and the assignment of genotypes and phenotypes should be called the *genotype-phenotype map*. The process of Darwinian selection is based on the *fitness* of phenotypes. This *valuation process* may be schematically represented by

$$\text{genotype} \longrightarrow \text{phenotype} \longrightarrow \text{fitness}$$

In fact all forms of life are determined by games based on sequences of amino acids which are valuated through the corresponding phenotype (Gatlin, 1972; Ratner, 1983; Volkenstein, 1975).

So far there is no complete model for any concrete biological system. This is due to the enormous difficulty of the biological valuation process. Special models for the evolution of prebiological systems were investigated e.g. by Eigen, Schuster (Eigen et.al., 1977, 1978; Eigen et.al., 1989), Anderson and Stein (Anderson, 1983; Anderson et.al., 1983).

This work is devoted to the investigation of games based on sequences. These games are characterized by a evolutionary dynamics based on certain artificial valuation process. As prototype we consider the frustrated game proposed by Engel (Ebeling, Engel, Feistel, 1990).

The idea of this paper is that the genotype-phenotype map transforms the rugged valuation landscape of fitness values of genotypes to a smooth fitness function of phenotypes and facilitate in this way the process of evolutionary search. On the other hand this may be interpreted as a mapping of an optimization problem to an intermediate level of coding, which is reflected in the representation problem of evolution"ary strategies, genetic algorithms and genetic programming (Rechenberg, 1973; Goldberg, 1989; Koza, 1992).

The fitness landscape proposed by Engel shows a rugged structure which is related to frustration of the problem (Ebeling et. al., 1994). We use the results of first three sections to characterize the valuation landscape. The main result is the determination of the density of states $n(F)$ by simulation and theoretical investigation.

Dynamical processes based on strings may be of some interest also for other fields of scientific activity: As we well know, the dynamics of information processing in human systems is based on the storage and exchange of the messages coded by strings of letters. Further we mention that many optimization processes, as e.g. the search of the travelling salesman, may be mapped on games with linear strings of letters. Finally we would like to point out that by the method of symbolic dynamics any trajectory of a dynamic system may be mapped on a string of letters on certain alphabet (Ebeling et.al., 1991).

## II. GENERAL PROPERTIES OF A FITNESS LANDSCAPE ON SEQUENCES

In the following we consider a set of $N$ sequences of length $L$, forming certain region in the sequence space. We assume that the elements of the sequence are taken from an alphabet consisting of $\lambda$ types of letters. The complete set of different sequences of length $L$ consists of $N_L = \lambda^L$ elements. For $L \gtrsim 100$ this number is astronomical. The most of possible sequences may be forbidden in realistic systems leaving only a subset of $N$ admitted ones for participation in the game.

Let us consider a mutation that takes place in a sequence. To measure the heaviness of the change we need a metric (distance) on the sequence space. The geometry of the space determines possible metric measures, we have to choose one of them. A standard metric may be introduced by means of the Hamming prescription. The Hamming distance between two sequences is defined as the number of non-coincidences.

We may define the *neighbourhood structure* of the sequence space due to this metric: Two sequences $s, s'$ with a hamming distance $h(s,s') = 1$ are neighbours. The neighbourhood structure is given by the adjacent matrix with $\mathcal{A}(s,s') = 1$ for neighbours and otherwise $\mathcal{A}(s,s') = 0$. A geometrical visualization of the neighbourhood structure may be given by a graph with edges connecting the neighbouring sequences. For $L = 1, 2$ with the alphabet $\{A, B, C, D\}$ this looks like

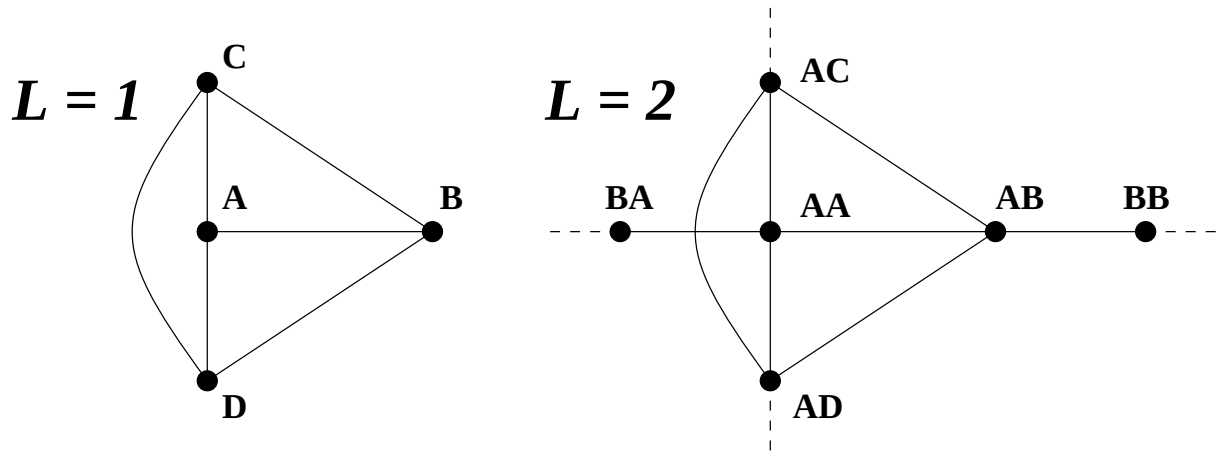

FIG. 1. *Neighbourhood structures of the sequence space $L = 1, L = 2$ due to the Hamming metric.*

In the case $L = 2$ we draw only one of the four identical components of the adjacent graph, all four components are strongly connected with each other. For $L \gtrsim 100$ the space has an extremely high number of points. The neighbourhood structure of the sequence space shows the feature, that all its points are near one to another, since the maximal number of non-coincidences is $L$ and the adjacent graph is very strongly connected.

Without loss of generality we consider the sequences as the *genotypes* of the individuals. All possible sequences may be considered as elements of a metric space, the genotype space $G$.

An individual comes into being by expression of a genotype $s \in G$. The individual has properties building its *phenotype* $\phi$, in this way we introduce also a phenotype space $Q$. The genotype-phenotype map $\Phi : s \in G \rightarrow \phi \in Q$ assigns the genotypes $s$ to phenotypes $\phi$. Each phenotype will be valuated in the selection process which determines the fitness function $F(\phi)$. When this happened we can say: The genotype $s$ was valuated with the fitness value $V(s) = F(\phi)$, $\phi = \Phi(s)$. The surface formed by the fitness values on the sequence space is called the valuation landscape (Conrad, 1983; Conrad et.al., 1992). Strictly speaking the landscape consists only of a discrete set of points, for imagination we connect these points by a surface, e.g. by a piecewise linear approximation.

The genotype-phenotype map $\Phi$ is the representation of genotypes in the valuation process which has built the valuation landscape $V$ by genotype expression of phenotypes and selection with $F$. The fitness value $V(s)$ may be an element of a real vector space. For simplification we confine ourself to the case that $V(s) \in \mathbb{R}_+$ is just a positive real number. Thus the valuation process is given by

$$\begin{array}{ccc} G & \xrightarrow{V} & \mathbb{R}_+ \\ {\scriptstyle\Phi}\searrow & & \nearrow{\scriptstyle F} \\ & Q & \end{array} \tag{1}$$

Let us introduce now the so-called density of states, a term borrowed from solid-state physics, as a first measure of the structure of the fitness landscape. We assume that the value is bounded from above and from below in the set of sequences $V_{min} < V_i < V_{max}$. We define the total number of sequences having the value $V_i < V$ by $N(V)$ and the relative occurrence by $S(V) = N(V)/N_L$ where $N_L$ is the total number of admitted sequences of length $L$. Here $N(V)$ and $S(V)$ are step functions converging to the values $N_L$ or 1 respectively.

We expect that the sequences having values in the interval $[V, V + dV]$ form a kind of density which we call the *density of states* $n(V)$. The density of states which formally is the derivative of $N(V)$ consists of delta-peaks. Correspondingly a normalized density of states $\sigma(V)$ may be derived from $S(V)$. Later on we shall use these concepts for a structural characterization of the landscape. We mention that the integral and differential number densities are invariant with respect to any ordering or choice of the neighbourhood structure on the sequence space.

## III. THE SMOOTHING REPRESENTATION AND THE GENOTYPE-PHENOTYPE MAPPING

At the beginning of a general investigation of evolution we have to ask the question: *Why does evolution work on sequence spaces?* In fact we now, evolution finds the extrema of a *smooth* fitness landscape very well. There exists a gradient way to extrema. The evolutionary dynamics is able to follow this way without sticking in local extrema. On rugged landscapes it is very difficult for evolution (and all other optimization strategies) to find a way to extrema (Kaufman, 1989, 1990, 1993). Consequently evolution had to establish a *smoothing representation* of the valuation landscape on sequence spaces for successful search.

On the other hand, there is the evidence that evolution does not valuate genotypes directly: The fitness function $F(\phi)$ values the phenotypes in the selection process and then the fitness values of genotypes $V(s) = F(\Phi(s))$ are only given by means of the genotype-phenotype map.

It is highly probable that the *smoothing representation* and the *genotype-phenotype map* are two interpretations of the same fact: The evolution had to choose the genotype-phenotype map in such manner that the representation of genotypes leads to a smooth fitness function. This is a necessary condition for efficient search in sequence spaces.

Indeed, it is very well known that the representation problem in evolution"ary strategies, genetic algorithms and genetic programming (Rechenberg, 1973; Goldberg, 1989; Koza, 1992) is the crucial point for success. Finding a representation is a complicated problem - there exists no algorithm to choose a good representation - it is only solvable with human inspiration. When a good representation was found evolutionary algorithms works very well. Exactly in this sense, evolution had to find a smoothing representation for genotypes and already the fact that we are able to think about it shows: it was found.

First let us define the meaning of a smooth representation of the fitness function on discrete spaces. We demand that the function is $\epsilon$-*continuous* in the following sense:

**Definition 1** *Let $F : Q \to I\!R_+$ be a function on a discrete space $Q$ with neighbourhood structure $\mathcal{A}$. We call $F$ $\epsilon$-continuous with the degree $\frac{\max F(s)}{\epsilon}$,*

$$\epsilon = \langle |V(s) - V(s')| \rangle_{s,s' \in G,\, \mathcal{A}(s,s')=1} \;. \tag{2}$$

*A function with a higher degree is more continuous.*

The phenotype should be some thing like the properties used by common sense to characterize the fitness of species (e.g. strength, robustness, speediness) and directly determine the fitness of an individual. A small alteration in the phenotype cause a small change in fitness. That is the *principle of strong causality* (Rechenberg, 1973). Thus we give following definition

**Definition 2** *The phenotype variables $\phi^\mu$ are parameters which are able to represent the fitness function $F : Q \to I\!R_+$ as a bijective function with a sufficient high degree of $\epsilon$-continuity on the phenotype space $Q = \{\phi = (\phi^1, ..., \phi^\mu)\}$.*

The $\epsilon$-continuity of $F$ guarantees that a small change $\Delta\phi$ corresponds to a small fitness difference $\Delta F(\phi)$. The bijectivity of $F$ provides that the fitness of an individual is unique determined by its phenotype and vice versa.

Now we asking for the question: It is possible to construct for a given problem a genotype-phenotype map $\phi = \Phi(s)$ in such a way, that for all possible genotype states $s$ and phenotype states $\phi$ the fitness function $F(\phi)$ can be given by definition 2? In general, can such a map $\Phi$ exists?

Every genotype state $s$ has the fitness value $V(s) = F(\Phi(s))$, but it is possible that many $s \in G$ have the same value $V$. On the other hand for each fitness value $F(\phi)$ there exists a unique phenotype state $\phi$. Thus we formulate the following theorem

**Theorem 1** *Two genotypes $s, s' \in G$ are equivalent by the relation*

$$s \sim s' \Leftrightarrow V(s) = V(s'). \tag{3}$$

*The phenotype states $\phi$ are the equivalence classes of the genotypes with respect to the valuation $V(s)$ of genotype states $s$ and fitness value $F$.*

$$\phi(F) = \{[s] : [s] \in G/\sim,\, V(s) = F\}. \tag{4}$$

*There exists a unique ordering procedure changing the neighbourhood structure $\mathcal{A}$ of $G/\sim$ in such a manner that the fitness function $F$ has a higher or equal degree as the valuation landscape $V$ with respect to $\epsilon$-continuity. From this, the genotype-phenotype map*

$$\Phi : G \to Q \qquad s \mapsto \phi \tag{5}$$

*can be uniquely determined by the ordering procedure. For the proof see appendix A.*

Let us explain the idea of the proof. The sequence space $G$ has the structure of a graph represented by the adjacent matrix $\mathcal{A}$. In general the building of equivalence classes leads to a change in the topology. The following picture illustrate this fact in this special case.

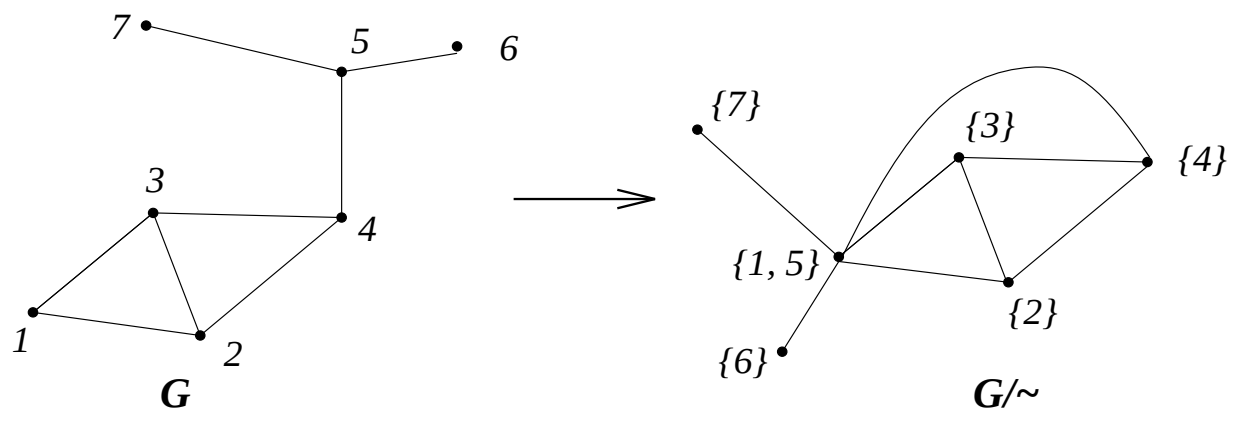

FIG. 2. *Neighbourhood structures of $G$ and $G/\sim$.*

The degree of continuity of $V$ does not change by this process. If we consider point mutations of sequences then these mutations form a group $\mathcal{G}$ generated by finite elements $g_1, \ldots, g_\nu$(generators). There is a finite number of point mutations transforming a sequence into an other one which has the same fitness value. If we represent the sequence by a point and every group generator by a line then the process of building equivalence classes can be described by the graph

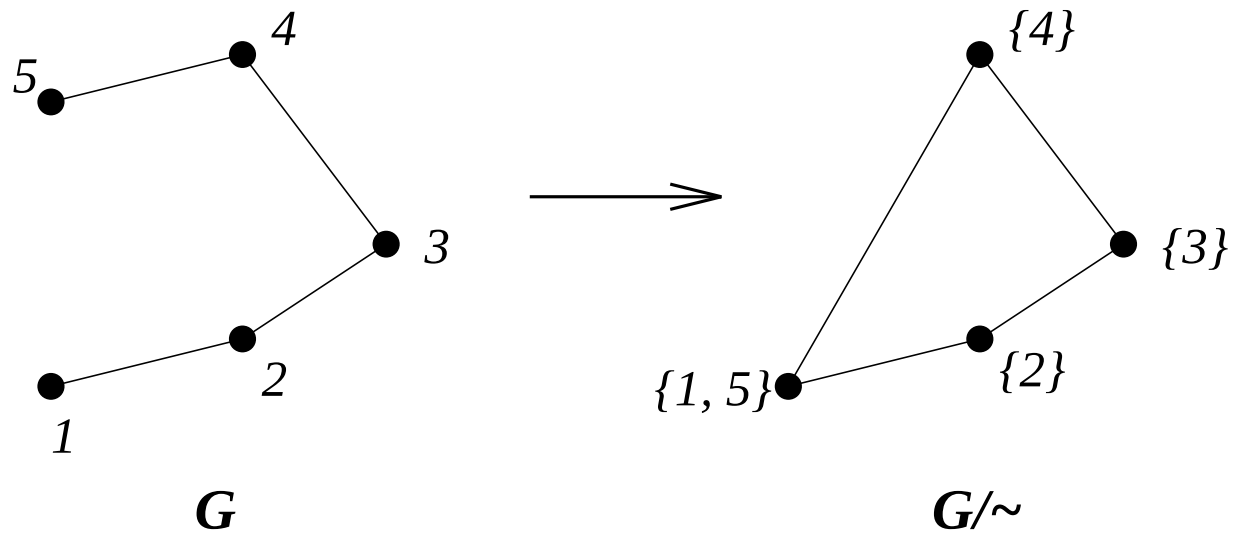

FIG. 3. *Action of the mutation on $G$ and on $G/\sim$*

So that the ambiguity comes from loops generated by the building of equivalence classes and represented by the mutation group $\mathcal{G}$. With two little rules (see appendix A) these loops can be eliminated to get a valuation with higher degree. By definition this is the property of the fitness function and we can identify the phenotype space with the space of equivalence classes together with the rules to smooth the landscapes. This defines by a unique procedure the genotype–phenotype map $\Phi$. Fig. 4 is a good example for this process.

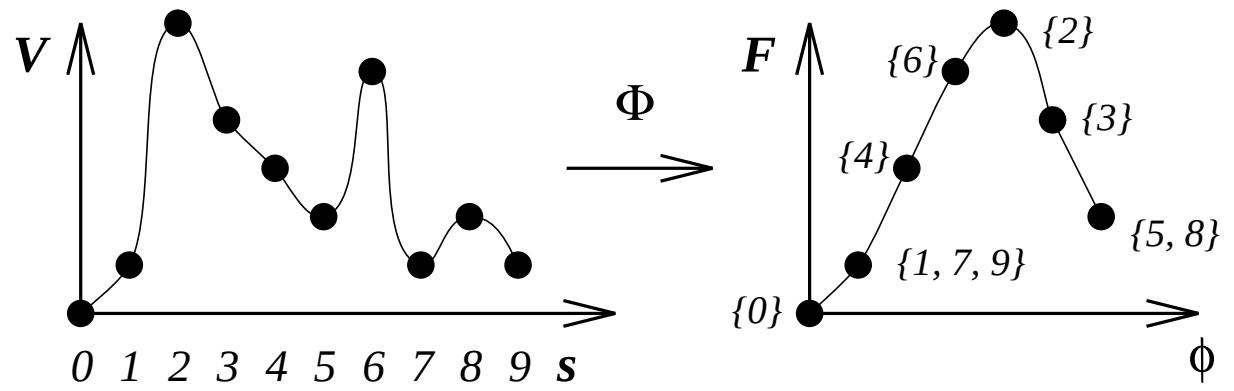

FIG. 4. *Genotype-phenotype map* $\Phi(s)$.

Obviously, the best genotype–phenotype map is the RANK-operator defined by ordering the fitness values to maximize the degree of the fitness function (fig. 5). The existence of the RANK-operator is an example of the ordering procedure introduced in theorem 1.

In section II we introduced the density of states $n(F)$. Now we can explain the meaning of this measure: Genotypes with the same fitness value build an equivalence class - the corresponding phenotype. The number of genotypes of a certain equivalence class (phenotype) is the density of states (fig. 5). Obviously, the density of states is only related to the ordering of fitness values but there are no references to the geometry and topology of the fitness landscape.

The density of states answers of the question: How difficult is it to find a certain phenotype? The density of states will be very low on hight fitness levels. Problems with a very fast slowdown of $n(F)$ will be very hard to solve. In this sense we may say: The density of states is a classifying measure of fitness landscapes. If we know $n(F)$ of two problems we are able to decide which problem is more difficult.

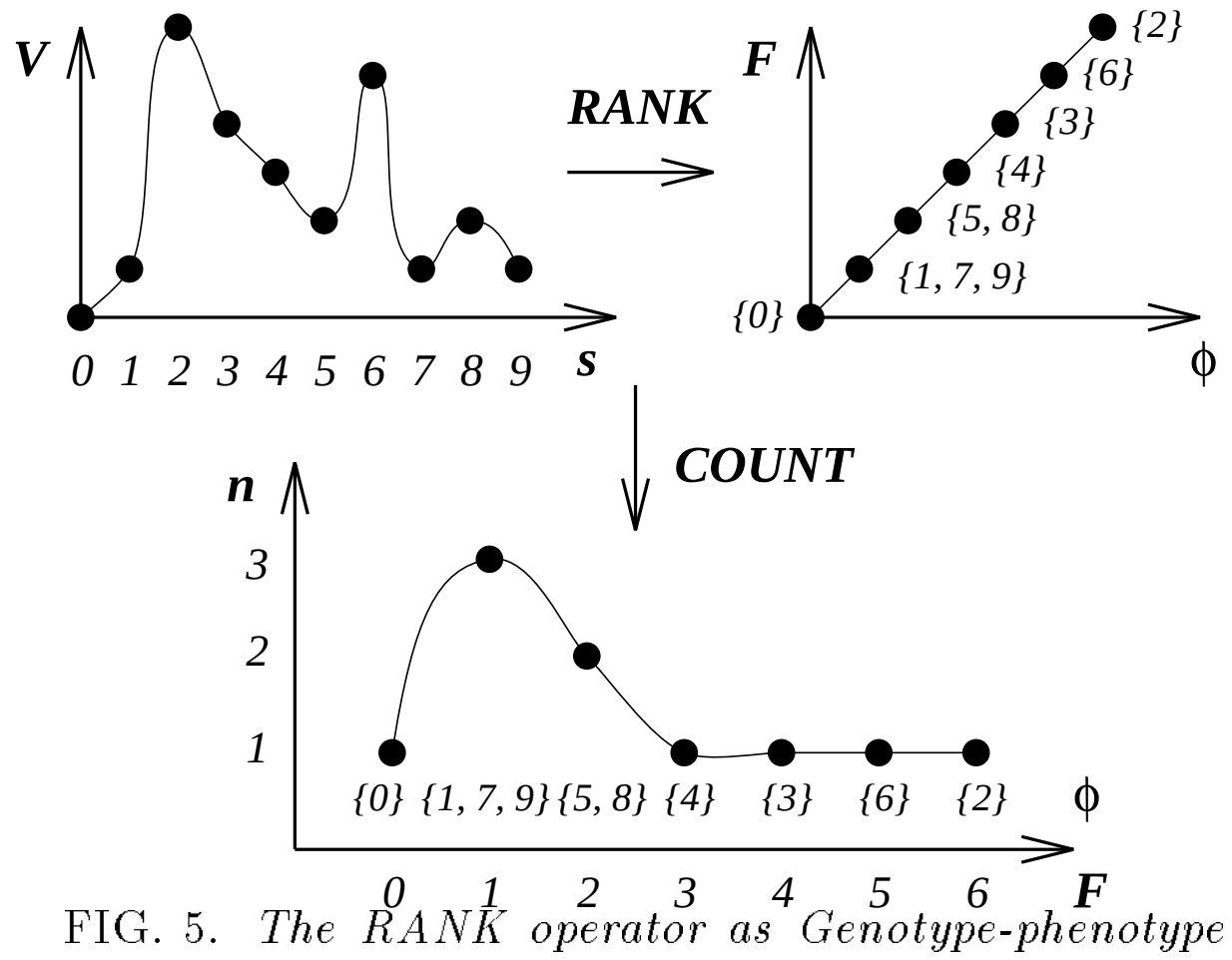

FIG. 5. *The RANK operator as Genotype-phenotype map and the density of states* $n(F)$.

## IV. TOY MODELS OF THE EVOLUTIONARY DYNAMICS

The evolutionary dynamics take place on spaces described in the previous sections. The valuations process of the evolutionary dynamics is characterized by a intricate genotype-phenotype map and a fitness function on the phenotype space. For simplicity of the introduction we confine ourselves to the case that the genotype and phenotype space are identical (The genotype-phenotype map is the identity).

We consider a genotype space of sequences $s \in G$ and choose a fixed numbering of the genotypes (Gödel number) $1, \ldots, i, \ldots, n$. The most simple model of an evolutionary dynamics is the Fisher-Eigen model which is based on the assumption that the competing objects $i = 1, \ldots, n$ have different reproduction rates $V_i$. These rates play now the role of the fitness. The evolutionary dynamics is given by the differential equations (Fisher, 1930; Eigen, 1971)

$$\dot{x}_i = (V_i - \langle V \rangle) x_i \, , \; \langle V \rangle = \sum_i V_i x_i \, , \; \sum_i x_i = 1 \, , \qquad (6)$$

where $x_i$ the fraction of individuals with the genotype $i$ in the population. The species with values better than the "social" average $\langle V \rangle$ will succeed in the competition and the others will fail. Finally only the species with the largest rate $V_{max}$ will survive.

In this way the Fisher-Eigen game explores the fitness landscape, finding out the peaks. The Fisher-Eigen model is the simplest of all models of competition. It refers to an oversimplified case and one may say that the model reflects only pseudo-competition since there is no real interaction between the species. Evolution needs not only competition but also mutations. We introduce mutation by an additional term in the dynamic equation

$$\dot{x}_i = \kappa (V_i - < V >) x_i + \sum_j [A_{ij} x_j - A_{ji} x_i] \qquad (7)$$

The scalar parameter $\kappa$ was introduced to allow a change of the strength of competition $\kappa$. One may consider three cases (Boseniuk et.al., 1987; Boseniuk et.al., 1990):

(i) Model of *Darwinian evolution*:

$$\kappa = 1; \quad A_{ij} = C_{ij}$$

where $C_{ij}$ is a symmetrical matrix $C_{ij} = C_{ji}$ of mutations. The symmetry of the mutation rates models the isotropy of biological mutations.

(ii) Model of *Boltzmann evolution* (Metropolis algorithm):

$$\kappa = 0, \; \mu = 1,$$
$$A_{ij} = C_{ij} \begin{cases} 1 & \text{if} \quad V_i > V \\ \exp[(V_i - V_j)/T] & \text{else} \end{cases} \; . \qquad (8)$$

Here the real positive parameter is the "temperature" of the Boltzmann search.

(iii) *Mixed Boltzmann-Darwin strategies*:
In this case we make the choice $0 \leq \kappa \leq 1$ and the Boltzmann-type mutation rate with $T > 0$.

We mention that the case (i) corresponds now to $\kappa = 1$ and $T = 0$. Further the case (ii) corresponds to $\kappa = 0$ and $T > 0$.

The basic elements of games playing the evolutionary dynamics are competitive self-reproduction which occurs with the rate $V_i$ and mutation which produces a genotype $i$ from $j$ with the rate $A_{ij}$. Selection is introduced by the condition of constant population size.

## V. A FITNESS FUNCTION WITH TWO FRUSTRATED PERIODICITIES

In the general case the valuation of a bio-sequence may be extremely complicated since it is based not on the primary structure itself but on a valuation of the corresponding phenotype. Here we restrict ourselves to an extremely simple model which was proposed by Engel in 1989 (Ebeling et.al., 1990). We also mention the similarity to 1D spin glasses with two different interaction ranges. We simplify the Engel-model by closing the sequences to rings. In this model the valuation $V(s)$ of the sequences $s$ over the alphabet $\{\mathbf{A},\ \mathbf{B},\ \mathbf{C},\ \mathbf{D}\}$ and the length $L$ is based on the following simple rules

(i) If a letter is in alphabetical order **ABCDA**... with the following letter then the fitness will increase by one, i.e $V = V + 1$.

(ii) If letters on position $i$ and $i + p$ are the same then $V = V + b$.

If the $i$-th element of a sequence $s$ is denoted by $S_i$ the valuation function (fitness value on the genotype space) is given by

$$V(s) = \sum_{i=1}^{L} [\alpha(S_i) + b\pi(S_i)] \tag{9}$$

where $\alpha(S_i) = 1$ if $S_i, S_{i+1}$ in alphabetical order and $\pi(S_i) = 1$ if $S_i = S_{i+p}$ otherwise $\alpha(S_i) = \pi(S_i) = 0$. The first rule (1) favours alphabetical sequences **ABCD-ABCDABC...D** with period 4. The second rule (2) favours periodic repetitions with the period $p$. If $p \neq 4$ then the tendencies to generate strings with period 4 or $p$ are contradictory, i.e. the system is frustrated. We choose $p = 5$ and set $b = b_c = 1/L\,[L/p]$ (Ebeling et.al., 1990), $b \ll b_c$ favours alphabetical sequences and $b \gg b_c$ $p$-periodic ones.

The valuation landscape $V$ has a rugged structure, i.e. sequences which are quite near with respect to their Hamming distance may have very different values. In the third section a general algorithms to smooth the fitness landscape was introduced. Now we want to present a example of this procedure in the case of the evolutionary game. Because of the linear structure of the fitness function the building of equivalence classes leads also to a linear structure. If we mutate an element $S_i$ of a sequence the change of fitness may be independent of the special letter of $S_i$, e.g. $\Delta V(\mathbf{AB} \to \mathbf{BB}) = \Delta V(\mathbf{AB} \to \mathbf{CB})$. Because of this fact we introduce a new description of the sequences which will allow us to find out the equivalence classes (phenotypes) of genotypes.

At first we looking for all possible transitions of sequences due to point mutations. A point mutation of the element $S_i$ of a sequence changes only the alphabetical orders $\alpha_i = \alpha(S_i)$, $\alpha_{i-1}$ and the $p$-periodicities $\pi_i$, $\pi_{i-p}$ (because of the direction in the arrangement of letters). The set of all possible transitions $(\alpha_{i-1}, \alpha_i)_t \longrightarrow (\alpha_{i-1}, \alpha_i)_{t+1}$ from time step $t$ to $t + 1$ is given by

$$\begin{aligned}
(00) &\longrightarrow \begin{cases} (00) = e \\ (01) = g_1 \\ (10) = g_2 \\ (11) = g_4 \end{cases} \\
(01) &\longrightarrow \begin{cases} (00) = g_1^{-1} \\ (10) = g_3 \end{cases} \\
(10) &\longrightarrow \begin{cases} (00) = g_2^{-1} \\ (01) = g_3^{-1} \end{cases} \\
(11) &\longrightarrow \{ (00) = g_4^{-1}
\end{aligned} \tag{10}$$

These transitions form a group $\mathcal{G}_\alpha$ generated by $\{e, g_1, g_2, g_3, g_4\}$ with respect to the relation $g_1 g_2^{-1} g_3 = e$ and the group operation is the concatenation of generators. That means $\mathcal{G}_\alpha$ has the structure of a free group.

For the $p$-periodicities $(\pi_{i-p}, \pi_i)$ one can easy found the same structure of transitions. Thus, the point mutation group $\mathcal{G}_\pi$ has the same structure like $\mathcal{G}_\alpha$, i.e.

$$\mathcal{G}_\alpha \simeq \mathcal{G}_\pi \tag{11}$$

both groups are isomorphic. The two states $\alpha_i$ and $\pi_i$ of the element $S_i$ define the *scheme state*

$$\mathfrak{f}_i = \begin{pmatrix} \pi_i \\ \alpha_i \end{pmatrix} = \begin{cases} \begin{pmatrix} 0 \\ 1 \end{pmatrix} = @ & : \text{ alphabetical} \\ \begin{pmatrix} 1 \\ 0 \end{pmatrix} = \# & : \ p-\text{periodic} \\ \begin{pmatrix} 1 \\ 1 \end{pmatrix} = \$ & : \text{ both} \\ \begin{pmatrix} 0 \\ 0 \end{pmatrix} = * & : \text{ none} \end{cases} \tag{12}$$

Together with the action of group elements on a scheme $\mathfrak{f} = \mathfrak{f}_1 \cdots \mathfrak{f}_i \cdots \mathfrak{f}_L$ we obtain the following fitness change for every point mutation $g \in \mathcal{G}_\alpha \times \mathcal{G}_\pi$

$$\triangle f(g) = f(\mathfrak{f}_{t+1}) - f(\mathfrak{f}_t)\,,\ \mathfrak{f}_{t+1} = g\,\mathfrak{f}_t \tag{13}$$

where the index $t$ denotes the time step. Thus, the possible changes of fitness by one point mutation $g = (g_\alpha, g_\pi)$ read as

$$\triangle f(g_\alpha) = \begin{cases} 0 : e \\ 1 : g_1 \\ 1 : g_2 \\ 0 : g_3 \\ 2 : g_4 \end{cases},\ \triangle f(g_\pi) = \begin{cases} 0 : e \\ b : g_1 \\ b : g_2 \\ 0 : g_3 \\ 2b : g_4 \end{cases} \tag{14}$$

with $\triangle f(g) = \triangle f(g_\alpha) + \triangle f(g_\pi)$, $\triangle f(g^{-1}) = -\triangle f(g)$.

The description of all possible transition by means of schemes $\int$ leads to a characterization of equivalence classes of sequences with respect to the fitness levels.

(i) *Class of interchangeable letters*:
We choose a new encoding of the letters {**A, B, C, D**} $\rightarrow$ {@, #, $, *} which transforms the sequence $s = S_1 ... S_L$ into the scheme $\int = \int_1 ... \int_L$, e.g. the sequences **BCDADAA** and **CDABABB** belong to the same class of the scheme $\int$ = @@@*$#@ with $V(\int) = 5 + 2b$. We can interchange the letters **A** $\rightarrow$ **B**, **B** $\rightarrow$ **C**, **C** $\rightarrow$ **D**, **D** $\rightarrow$ **A** without changing the scheme and fitness value.

(ii) *Class of permutable schemes*:
The schemes @@@*$#@ and @@@@*$# are the same up to one translation. To characterize the classes of schemes which differ only by translation and permutation we encode the scheme @@@*$#@ $\rightarrow (1, 1, 4, 1)$ by the *scheme vector* counting the numbers of ($, #, @, *) in the scheme. E.g. for $b = 0.1$ and $L = 7$ we can find

| $V$ | scheme | $ | # | @ | * |
|---|---|---|---|---|---|
| 6.0 | @@@@@*@ bcdabca | 0 | 0 | 6 | 1 |
| 5.2 | *$$@@*@ babcdaa | 2 | 0 | 3 | 2 |
| | @@@*$#@ bcdadaa | 1 | 1 | 4 | 1 |

(iii) *Class of fitness levels*:
The scheme vectors $(1, 1, 4, 1)$ and $(2, 0, 3, 2)$ have the same fitness $V = 5 + 2b$. We build the equivalence classes of fitness levels by counting the numbers of alphabetical and $p$-periodic letters $(\alpha, \pi)$, e.g. $(1, 1, 4, 1) \rightarrow (5, 2), (2, 0, 3, 2) \rightarrow (5, 2)$. The state $(\alpha, \pi)$ determines unique the fitness value $V(\alpha, \pi) = \alpha + b\,\pi$. Thus we call the numbers $(\alpha, \pi)$ of alphabetical and $p$-periodic letters of a sequence $s$: the phenotype $\phi$.

Thus, the genotype-phenotype map of the system is given by

$$\Phi : s \in G \rightarrow \phi = (\alpha, \pi) \in Q \,, \qquad (15)$$
$$\alpha = \sum_{i=1}^{L} \alpha(S_i) \;\;,\;\; \pi = \sum_{i=1}^{L} \pi(S_i) \;.$$

We emphasize not all combinations of $\alpha$, $\pi$ are possible. The structure of $\Phi$ is just determined by this restrictions.

Now, we are able to show the smoothing action of the genotype-phenotype map. We consider sequences of the length $L = 7$ and $b = 0.1$ and choose a *Hamilton way* through the genotype space $G$. That means, we give every sequence $s_i \in G$ a Gödel number $i$ in such a manner that $s_i$ and $s_{i+1}$ are neighbours due to the Hamming metric of $G$. Fig. 6 shows the fitness values of the sequences numbered due to the Hamilton way in a representative range. It is easy to see that the valuation landscape has a very rugged structure, the degree of $\epsilon$-continuity is 1.26.

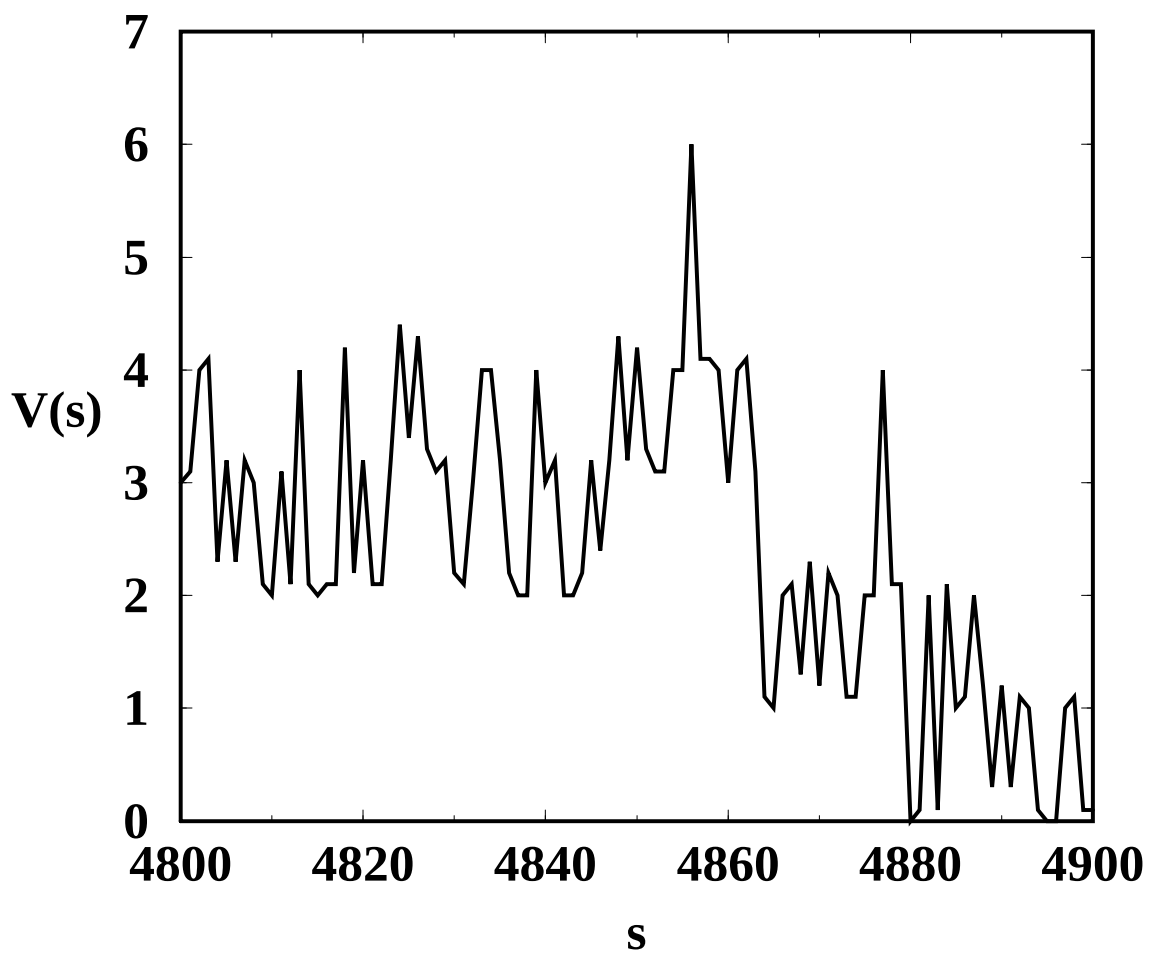


FIG. 6. *The valuation landscape on the genotype space, $L = 7$.*

The fitness landscape on the phenotype space $F(\alpha, \pi)$ is shown in fig. 7. We can see the landscape over $(\alpha, \pi)$ is very smooth. Not all combinations $(\alpha, \pi)$ are possible phenotypes, the optimum $F(6, 0) = 6.0$ is an isolated island on the landscape (right side fig. 7). The degree of $\epsilon$-continuity is 0.379.

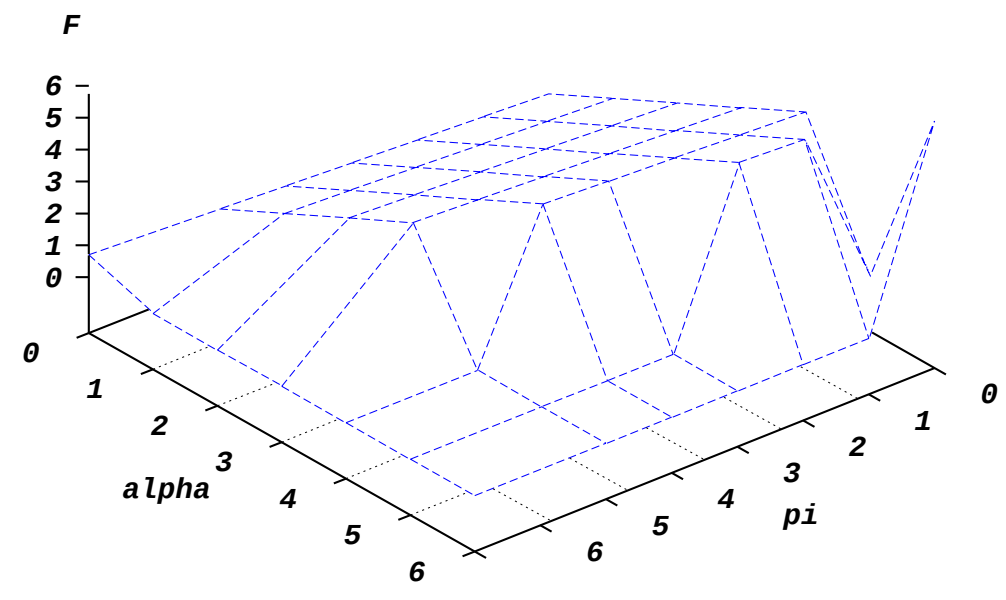


FIG. 7. *The fitness landscape on the phenotype space, $L = 7$.*

The figures clearly show the smoothing action of the genotype-phenotype map.

The density of states has been defined by the number of genotypes belong to a certain phenotype. Fig. 8 shows the density over the phenotypes $n(\alpha, \pi)$ and the fitness levels $n(F)$. On the one hand side, the density of states seems to be a very rugged function when we looking at $n(F)$. On the other hand, the density over the phenotypes $\phi = (\alpha, \pi)$ is a very smooth landscape. This interesting feature underlines the importance of the right choice of the genotype-phenotype map: The phenotypes

obtained by (15) seem to be the *natural* representation of the problem.

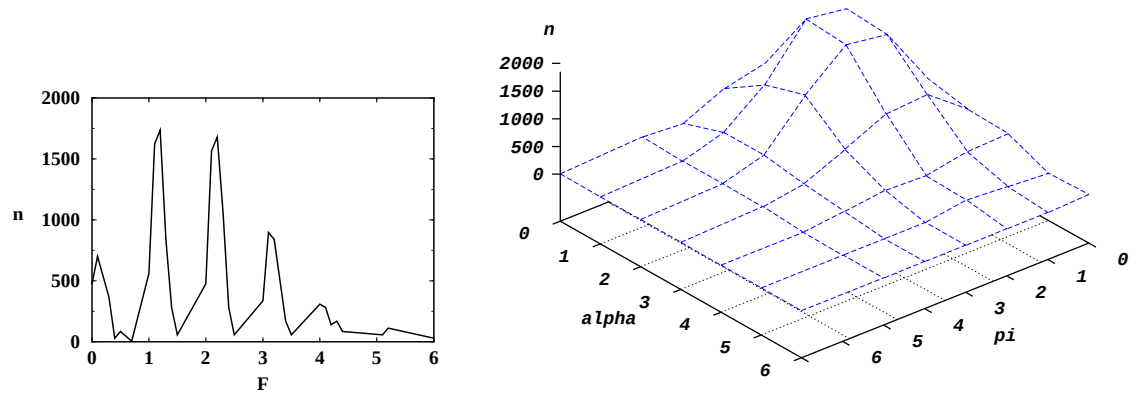


FIG. 8. *The density of states $n(F)$ and $n(\alpha,\pi)$, $L=7$.*

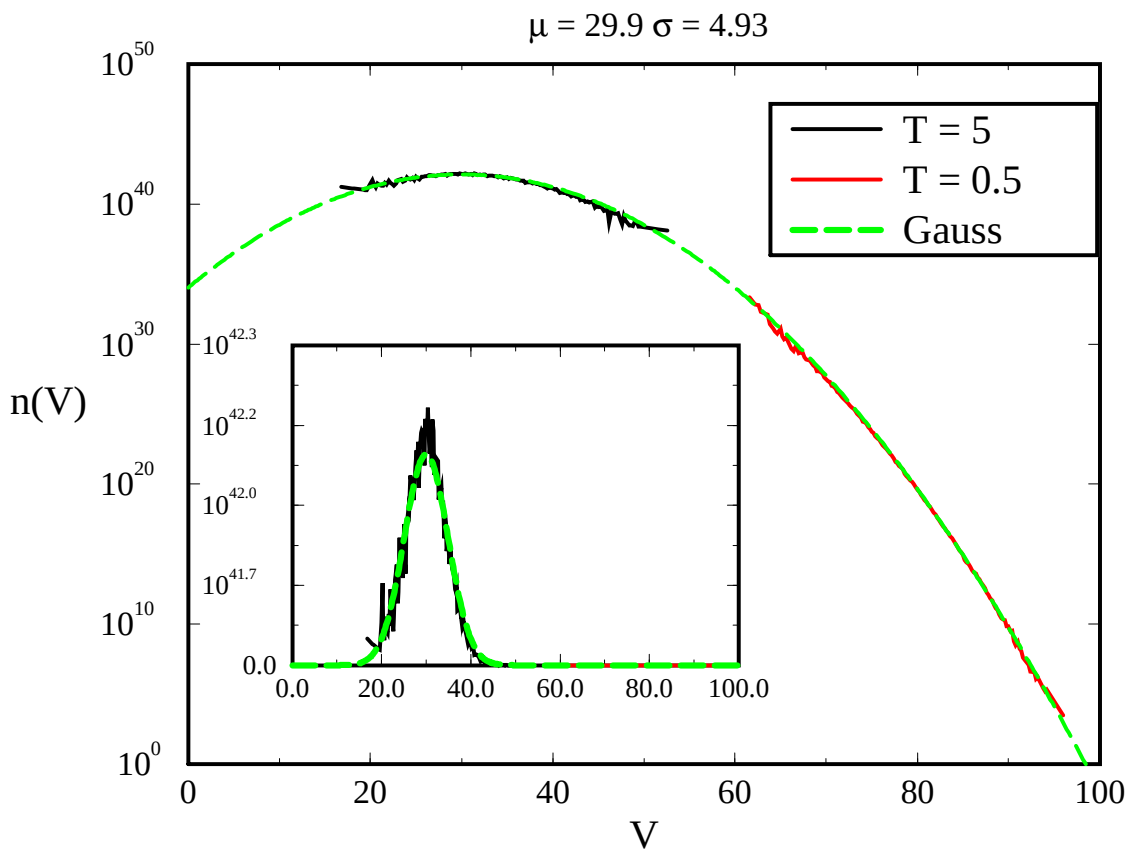


FIG. 9. *The density of states $n(F)$, $L=100$.*

## VI. THE DENSITY OF STATES OF LONG SEQUENCES

The number of possible sequences with $L=100$ is approximately $10^{60}$, thus the calculation of $n(\alpha,\pi)$ or $n(F)$ is impossible. Fortunately, it is well known from statistical physics and the theory of thermodynamic strategies (Andresen, 1989; Sibiani et. al. 1990; Berry et.al., 1993) that for a Boltzmann strategy (Metropolis algorithm) the equilibrium density of realizations of values observes the canonical distribution

$$P_{eq}(V) \sim n(V) \cdot \exp\left(-\frac{V}{T}\right)$$

The density of states is given by

$$n(V) \sim P_{eq}(V) \cdot \exp\left(\frac{V}{T}\right) \qquad (16)$$

To obtain the equilibrium density $P_{eq}(V)$ we simulated an ensemble of $N=10,000$ sequences of length $L=100$ which carry out a Boltzmann strategy with the mutation rate (8) and the potential (9). The density $P(V)$ is approximated by the frequency $N(V)/N$, where $N(V)$ is the number of individuals with $V \in [V, V+\Delta V)$. In the long time limit $P(V)$ tends towards the equilibrium density $P_{eq}(V)$. After $10,000$ time steps $P(V)$ was relaxated into the equilibrium. We tested the convergency behaviour up to $100,000$ time steps.

We have made the simulation at two different temperatures to scan up the whole range of $V$. The resulting density of states is in very good approximation a Gaußdistribution as to be seen from fig. 9.

We emphasize that the method also works for other fitness landscapes and is not restricted to sequences. The simulation of ensemble of *Boltzmann searchers* is a general way to obtain the density of states and therefor a method to classify the fitness landscape of any optimization problem.

In the special case of a fitness landscape like (9) we are able to calculate the structure of $n(V)$ by means of the group $\mathcal{G}$.

Let $\mathcal{S}$ be the space of schemes and let $V : G \longrightarrow \mathbb{R}$ be the valuation function where $G$ is the sequence space. Now we introduce a equivalence relation by

$$s_1 \sim s_2 \Longleftrightarrow V(s_1) = V(s_2) \quad \forall s_1, s_2 \in G$$

So the set of equivalence classes $G/\sim$ is up to a little set of combinatorial operations equal to the space of scheme states $\mathcal{S}$. We are interested on the question: How many sequences with the same fitness value exist? To this end we introduce a map $n : \mathbb{R} \longrightarrow \mathbb{R}$ which gives for every fitness value the number of states occupying this value. We argue that for large sequence lengths $L$ the number of combinatorial operations for every equivalence class is constant.

Consider now a little shift of the valuation $V \longrightarrow V+\Delta V$. This shift can be expressed by group action of $\mathcal{G}$ which is defined in the following sense. The group relations

$$g_i g_i^{-1} = e \qquad i = 1,2,3,4 \quad \text{and} \quad g_1 g_2^{-1} g_3 = e \qquad (17)$$

in $\mathcal{G}$ are defined local that means for all places in the string. Now we define the group action by action of every generator on the sequence with valuation 0 with respect to the relations (17). Because of correlations we obtain the following restrictions for two generators acting on two successive places in the string

$$\begin{array}{cc} (g_3)_{i-1}(g_2)_i & (g_3)_{i-1}(g_4)_i \\ (g_3^{-1})_{i+1}(g_1)_i & (g_3^{-1})_{i+1}(g_4)_i \end{array}$$

where $(g)_i$ means the action of the generator $g$ on the $i$-th place. The action on the $i$ and $i+2$-th places are

independent. The next step is the formal introduction of derivations over the sequence space via group actions. Since Conne's non-commutative geometry many physicists applicate such methods to the case of discrete sets (Müller-Hoisen et.al., 1993). We define the derivative of $V$ along in the $i$-th direction of the scheme space by

$$\frac{\partial V}{\partial g}(s) = \frac{1}{L}(V(gs) - V(s)) = \frac{1}{L}(L_g - 1)V(s) \qquad s \in \mathcal{S} \tag{18}$$

where $L_g$ is the links translation given by $L_g V(s) = V(gs)$. From the ordinary rules we obtain

$$\frac{dn}{dV} = \sum_{g\in\mathcal{G}} \frac{\partial g}{\partial V}\frac{\partial n}{\partial g} \tag{19}$$

If we introduce the function

$$\delta(x - x_0) = \begin{cases} 1 & x = x_0 \\ 0 & \text{else} \end{cases} \tag{20}$$

then it follows

$$n(V(s)) = \sum_{g\in\mathcal{G}} \delta(V(gs) - V(s)) = \sum_{g\in\mathcal{G}} \delta((L_g - 1)V(s)) \tag{21}$$

For instance this expression can be approximated by

$$\sum_{g\in\mathcal{G}} \delta(V(gs) - V(s)) \approx \sum_{g\in\mathcal{G}} \exp(-B\cdot(V(gs)) - V(s))^2) \tag{22}$$

with $B \gtrsim L$ as suitable number. Let $h : \mathbb{R} \longrightarrow \mathbb{R}$ be an arbitrary function without singularity in $h(x_0)$ then

$$h(x)\delta(x - x_0) = h(x_0) \tag{23}$$

$$\frac{d}{dx}h(x)\delta(x - x_0) + h(x)\frac{d}{dx}\delta(x - x_0) = 0 \tag{24}$$

Together with (24) we obtain

$$\frac{dn}{dV} = -\sum_{g\in\mathcal{G}} \frac{h'(V)}{h(V)}\,\delta(V(gs) - V(s)) \tag{25}$$

Comparing with (19) and together with the obvious relation

$$\frac{\partial n}{\partial g}(g_1) = \delta(V(g_1 s) - V(s)) \tag{26}$$

it follows

$$-\frac{\partial g}{\partial V}(V) = \frac{h'(V)}{h(V)} = \frac{d}{dV}\ln(h(V)) \tag{27}$$

This formula can be interpreted as "density of generators acting on the scheme state with valuation $V$". Next we will calculate these expression by arguments relating to the structure of the group $\mathcal{G}$. The action of the generators $g_1$ and $g_2$ leads to increasing of the valuation by 1 or $b$, respectively. The probability to make such mutation is twice as big as the probability of the mutation with $g_4$. This follows simple from the relations (17). Because of the linear defined valuation we obtain

$$\frac{\partial g}{\partial V}(V) \sim V \tag{28}$$

If the valuation increase the density of generators will decrease because the number of states $*$ decrease after every mutation. That means

$$\frac{\partial g}{\partial V}(V) = A(V_0 - V) \tag{29}$$

with suitable constants $V_0, A$. Together with (25) and (27) the differential equation

$$\frac{dn}{dV} = -A(V - V_0)n \tag{30}$$

is obtained. The solution of this equation (for large $L$) is simple

$$n(V) = n(0)\exp(-A(V^2 - V_0 V)) \tag{31}$$

$$= n(0)\exp(AV_0^2/4)\exp(-A(V - V_0/2)^2) \tag{32}$$

So that by investigation of such methods the qualitative structure of the degeneration distribution is obtained. The proof of all formulas is mathematical not rigourous but this should simple done following (Müller-Hoisen et.al., 1993).

## VII. CONCLUSIONS

We have shown that the investigation of evolution on discrete spaces lead to a number of principal problems, e.g.:

(i) The topological properties of the genotype and phenotype space.

(ii) The structure of the fitness landscape.

(iii) The smoothness of the fitness landscape and the representation by the genotype-phenotype map.

(iv) The classification of fitness landscape and the derivation of optimal search strategies.

The question of the structure of the fitness landscape and its classification is connected with the topological properties of the underlying spaces. We have shown that the choice of the Hamming metric on the genotype space leads to a rugged valuation landscape on which the evolutionary search is very difficult. On the other hand it is possible to smooth the landscape by a suitable representation of the problem. The genotype-phenotype map

transforms the genotype space and its metric to the phenotype space and a new metric. On this space the principle of strong causality is valid: the change of fitness between two neighbouring phenotypes, with respect to the new metric, is very small. The genotype-phenotype map increased the degree of $\epsilon$-continuity of the landscape. The introduction of the new metric on the phenotype space may also described by new mutation operators on the genotype space: The suitable mutation operators transform a genotype in such a manner that the corresponding phenotypes are neighbours with respect to the new metric on the phenotype space. The determination of the genotype-phenotype map or the construction of new mutation operators are two interpretations of the same problem: the problem of a smooth representation of the fitness landscape.

The density of states is an measure of the difficulty of an optimization problem. This measure is invariant to the genotype-phenotype map or any choice of representation of the fitness landscape and characterized only the problem. That makes it feasible to use the density of states as classifying measure of fitness landscapes.

## APPENDIX A

**Lemma 1.**

**Proof 1** *At first we want to fix the topology in the sequence space given by the neighbourhood structure* $\mathcal{A}_G$. *The neighbourhood of a special sequence* $s \in G$ *will be denoted by* $N_G(s)$. *Next we introduce the equivalence relation*

$$s_1 \sim s_2 \Longleftrightarrow V(s_1) = V(s_2) \qquad \forall s_1, s_2 \in G$$

*and form the quotient space* $G/\sim$. $G$ *carry a natural semigroup structure given by concatenation of let-*

*ters. This structure can be extended to $G/\sim$. In general the building of equivalence classes leads to a change in the topology. Consider two sequences $s, s' \in G$ with $\mathcal{A}_G(s', s) = 0$ and $V(s) = V(s')$. Both sequences are representants of the same equivalence class. In the neighbourhood of each sequence are by definition sequences with differ only by one letter. It is obvious that this fact change the metric and the topology. Let $\mathcal{A}_\sim$ be the neighbourhood structure in $G/\sim$ induced from $\mathcal{A}_G$ in $G$. Consider three sequences $s, s', s'' \in G$ with $s \sim s'$ and $\mathcal{A}_G(s, s'') = 1$ ($s''$ is in the neighbourhood of $s$) so we obtain simple*

$$\mathcal{A}_\sim(s, s') = 0 \quad \mathcal{A}_\sim(s, s'') = 1 \quad \mathcal{A}_\sim(s', s'') = 1$$

*So it is obvious that the valuation function $\hat{V}$ over $G/\sim$ induced from the valuation of $G$ has the same $\epsilon$ number, that means*

$$\max_{\substack{s,s' \in G \\ \mathcal{A}_\sim(s,s')=1}} \left(|V(s) - V(s')|\right) = \max_{\substack{s,s' \in G/\sim \\ \mathcal{A}_\sim(s,s')=1}} \left(|\hat{V}(s) - \hat{V}(s')|\right) \tag{A1}$$

*But the function $\hat{V}$ differs from $V$ by the fact that $\hat{V}$ is bijective. The ambiguity of the valuation function $V$ is encoded in the topological structure of $G/\sim$. To proof this assertion we consider point mutation of the sequence. The set of point mutation forms a group denoted by $\mathcal{G}$ which together with a group action $a : \mathcal{G} \times G \longrightarrow G$ determines the point mutations completely. This group is generated by a finite number of elements $g_1, \ldots, g_\nu$ (generators) which are equal to the elementary mutations. Because of the existence of different sequences with the same valuation a finite number $k$ of mutations represented by a sequence of generators $g_{i_1} g_{i_2} \cdots g_{i_k}$ exists with the following property*

$$s \sim a(g_{i_1} g_{i_2} \cdots g_{i_k}, s) \qquad s \in G$$

*So that the ambiguity comes from loops generated by the building of equivalence classes and represented by the mutation group $\mathcal{G}$. To eliminate this ambiguity we have to change the topology of $G/\sim$ without changing the bijective map $\hat{V}$. This can be done by the following rules:*

*(i) Cut the line in the loop which has the represent the largest change in the valuation.*

*(ii) Connect the disjoint parts of the space generating by the rule (i) in such way that the change in the valuation will be minimized.*

*The rules generate a connected space denoted by $G_s/\sim$. The valuation function $\hat{V}$ does not change by this procedure and is denoted by $V_s$. Because of this fact we obtain from rules above that*

$$\max_{\substack{s,s' \in G \\ h(s,s')=1}} \left(|V(s) - V(s')|\right) \geq \max_{\substack{s,s' \in G_s/\sim \\ f_s(s,s')=1}} \left(|V_s(s) - V_s(s')|\right) \tag{A2}$$

*with $f_s$ as metric in $G_s/\sim$. The very nice fact is obtained that the valuation function $V_s$ is more smooth as $V$. If we denote the map from $G$ to $G_s/\sim$ by $p_s$ then the commuting diagram follows*

$$\begin{array}{ccccc} & & G_s/\sim & & \\ & {}^{p_s}\nearrow & & \searrow^{V_s} & \\ G & & \xrightarrow{V} & & \mathbb{R}_+ \\ & {}_{\Phi}\searrow & & \nearrow_{F} & \\ & & Q & & \end{array}$$

*i.e. $V_s \circ p_s = V = F \circ \Phi$. Because of the bijective functions $V_s$ and $F$ the genotype–phenotype map $\Phi = F^{-1} \circ V_s \circ p_s$ is completely characterized by $p_s$ and has the properties according to the theorem 1. Thus the genotype–phenotype map is uniquely given by the construction defined above.* **q.e.d.**